\documentclass[reprint,prl]{revtex4-1}
\usepackage{amsmath}
\usepackage{amssymb}
\usepackage{amsfonts}

\usepackage[T1]{fontenc}
\usepackage{bm}
\usepackage{color}
\usepackage{graphicx}
\usepackage[breaklinks=true]{hyperref}
\hypersetup{
    bookmarks=true,         
    unicode=false,          
    pdftoolbar=true,        
    pdfmenubar=true,        
    pdffitwindow=false,     
    pdfstartview={FitH},    
    pdftitle={My title},    
    pdfauthor={Author},     
    pdfsubject={Subject},   
    pdfcreator={Creator},   
    pdfproducer={Producer}, 
    pdfkeywords={keyword1} {key2} {key3}, 
    pdfnewwindow=true,      
    colorlinks=true,        
    linkcolor=blue,         
    citecolor=blue,         
    filecolor=blue,         
    urlcolor=blue           
}

\begin{document}

\title{Action principle for Coulomb collisions in plasmas}
\author{Eero Hirvijoki}
\email{eero.hirvijoki@chalmers.se}
\affiliation{Department of Applied Physics, Chalmers University of Technology, 41296 Gothenburg, Sweden}
\date{\today}

\begin{abstract}
In this letter we derive an action principle for Coulomb collisions in plasmas. Although no natural Lagrangian exists for the Landau-Fokker-Planck equation, an Eulerian variational formulation is found considering the system of partial differential equations that couple the distribution function and the Rosenbluth potentials. Exact conservation laws are derived after generalizing the energy-momentum stress tensor for second order Lagrangians and, in the case of a test-particle population in a given plasma background, the action principle is shown to correspond to the Langevin equation for individual particles. Being suitable for discretization, the presented action allows construction of variational integrators. Numerical implementation is left for a future study. 
\end{abstract}

\maketitle 

{\bf {\em Introduction}} -- 
The principle of stationary action and the Noether theorem provide powerful tools for analyzing mechanics of particles and fields. Using them, it is possible to find a conservation law for any symmetry in a Lagrangian. Both tools also apply to discrete mechanics, allowing one to construct variational integrators (see e.g.~\cite{marsden_west:2001,stern:2015}) which typically conserve the discrete counterparts of the conserved continuous quantities. It is no wonder that physicists have had a long-lasting love affair with action principles.

In plasma physics, simulations are entering an era where conservation of energy and momentum is not only a desired feature but a necessity: it would be difficult to trust the results of, for example, global turbulence simulations if numerical methods resulted in artificial damping or growth of turbulence. Superior numerical properties in mind, variational algorithms have been recently developed for particle dynamics~\cite{qin:2008:PRL,zhang:2015:POP,lelandellison:ppcf:2015} as well as for Vlasov-Maxwell and Vlasov-Poisson systems~\cite{squire_PoP_2012,kraus:PHD}. 

Action principles in plasma physics (see e.g~\cite{low_1958,ye_PoF_1992,larsson_JPP_1993,fla_PoP_1994,brizard_PRL_2000} for Eulerian formulations of the Vlasov-Maxwell system) are, however, still lacking the collisional effects. Emerging in the form of a collision operator in the kinetic equation for a single-particle distribution function, they describe the binary Coulomb interactions between particles and play an essential role providing equilibration and increasing entropy. Only together, the Vlasov-Maxwell system and the collision operator provide a fundamental description for plasmas.  An action principle for Coulomb collisions is thus desired and, in this letter, we provide one.

The Landau-Fokker-Planck equation, describing the collisional evolution of the distribution function in the velocity space, is a non-linear integro-differential equation and, unfortunately, does not exhibit a natural Lagrangian. Thanks to Rosenbluth~\cite{rosenbluth:PhysRev.107.1}, the Fokker-Planck equation can, however, be formulated as a set of coupled partial differential equations: a diffusion-advection equation for the distribution function $f$ in the velocity space, and two Poisson equations for potential functions $\phi$ and $\psi$ from which the velocity-space advection and diffusion coefficients can be computed. As Ibragimov~\cite{ibragimov_JMAA_2006,ibragimov_JMAA_2007} has found a way to obtain Lagrangians and conservation laws for any system of partial differential equations of arbitrary order, we combine the two ideas and find an action principle for Coulomb collisions. 

The letter proceeds by first shortly reviewing the classical field theory and its extension by Ibragimov for our purposes. We then derive a generalization of the energy-momentum stress tensor for Lagrangians containing derivatives up to second order, and apply the theory to the Rosenbluth-Fokker-Planck system. We obtain two second order Lagrangians and show that the total energy and momentum carried by the fields of the extended systems are conserved. Finally, before summarizing our results, we discuss the connection of our action principle to the Langevin equation for an individual particle. 

As variational integrators have already been constructed for specific nonvariational partial differential equations that admit first order extended Lagrangians~\cite{kraus:2015}, we anticipate the procedure to be applicable to second order Lagrangians, and to the Rosenbluth-Fokker-Planck system as well. Finding such a variational algorithm is, however, left for a future study.

{\bf {\em Mathematical preliminaries}} --
Throughout this letter, we use the following notation: the independent variables (coordinates) in the phase-space-time are denoted~by $x=\{x^{\mu}\}_{\mu=0}^{m}=(t,x^1,\dots,x^m)$ with $m$ the dimension of the phase-space and the phase-space components being $\{x^i\}_{i=1}^{m}$. The dependent variables (fields) are denoted by $u=\{u^{\alpha}\}_{\alpha=1}^{n}=(u^1,\dots,u^n)$ with $n$ the number of fields. Partial derivatives of the dependent variables with respect to the independent coordinates are denoted by the convention $u^{\alpha}_{\mu}=\partial u^{\alpha}/\partial x^{\mu}$, which allows for a compact notation for a collection of derivatives according to $u_{(1)}=\{u^{\alpha}_{\mu}\}_{\mu,\alpha}$, $u_{(2)}=\{u^{\alpha}_{\mu\nu}\}_{\mu\nu,\alpha}$ etc. We also use the differential
\begin{equation}
D_{\mu}=\frac{\partial}{\partial x_{\mu}}+u^{\alpha}_{\mu}\frac{\partial}{\partial u^{\alpha}}+u^{\alpha}_{\mu\nu}\frac{\partial}{\partial u^{\alpha}_{\nu}}+\dots,
\end{equation}
and define the action ${\cal A}[u]$ for the fields $u$ as the integral
\begin{equation}
{\cal A}[u]=\int {\cal L}(x,u,u_{(1)},u_{(2)},\dots)\, d x,
\end{equation}
where ${\cal L}(x,u,u_{(1)},u_{(2)},\dots)$ is the Lagrangian (density). Summation over repeated indices is assumed.

In most of the classical problems in physics that possess known Lagrangians, the Lagrangian contains derivatives only up to first order, i.e., ${\cal L}(x,u,u_{(1)})$, and the Euler-Lagrange equations resulting from the principle of stationary action take a familiar form
\begin{equation}
\frac{\partial {\cal L}}{\partial u^{\alpha}}-D_{\mu}\frac{\partial {\cal L}}{\partial u^{\alpha}_{\mu}}=0.
\end{equation}
In this case, the Noether theorem provides conservation laws for each symmetry in the Lagrangian according to
\begin{equation}
D_{\mu}j^{\mu}=0,\qquad j^{\mu}=\xi^{\mu}{\cal L}+(\eta^{\alpha}-\xi^{\nu}u_{\nu}^{\alpha})\frac{\partial{\cal L}}{\partial u^{\alpha}_{\mu}},
\end{equation}
where the flux $j^{\mu}$ is also known as the Noether current. The vector fields $\xi$ and $\eta$ appearing in $j^{\mu}$ have to satisfy the invariance condition
\begin{equation}
\label{eq:first_order_invariance_condition}
X({\cal L})+{\cal L}D_{\mu}(\xi^{\mu})=0,
\end{equation}
where the generator for the infinitesimal transformation, under which the Lagrangian is invariant, is given by 
\begin{equation}
X=\xi^{\mu}\frac{\partial}{\partial x_{\mu}}+\eta^{\alpha}\frac{\partial}{\partial u^{\alpha}}+\left[D_{\mu}(\eta^{\alpha})-u_{\nu}^{\alpha}D_{\mu}(\xi^{\nu})\right]\frac{\partial}{\partial u_{\mu}^{\alpha}}.
\end{equation}

According to Ibragimov~\cite{ibragimov_JMAA_2006,ibragimov_JMAA_2007}, one may construct a Lagrangian for an arbitrary system of partial differential equations of any order $(s)$ as long as the number of dependent fields equals the number of differential equations. The recipe is simple: for a set of $n$ differential equations for $n$ fields $f=\{f^{\alpha}\}_{\alpha=1}^{n}$ that are defined implicitly by
\begin{equation}
F_{\alpha}(x,f,f_{(1)},\dots,f_{(s)})=0, \quad \alpha=1,\dots,n
\end{equation}
the Lagrangian is given by
\begin{equation}
{\cal L}=f^{\star\alpha}F_{\alpha},
\end{equation}
where $f^{\star}=\{f^{\star\alpha}\}_{\alpha=1}^{n}$ are the so-called {\em auxiliary} fields. Further, the {\em adjoint} equations, that describe the evolution of the auxiliary fields, are given by 
\begin{equation}
F^{\star}_{\alpha}(x,f,f^{\star},f_{(1)},f^{\star}_{(1)},\dots)=\frac{\delta{\cal L}}{\delta f^{\alpha}}=0,
\end{equation}
where the {\em Euler-Lagrange operator} is defined by
\begin{equation}
\label{eq:euler-lagrange}
\frac{\delta}{\delta u^{\alpha}}\equiv\frac{\partial}{\partial u^{\alpha}}+\sum_{j}(-1)^jD_{\mu_1}\dots D_{\mu_j}\frac{\partial }{\partial u^{\alpha}_{\mu_1\dots\mu_j}}.
\end{equation}
In other words, the adjoint equations are found by requiring the action ${\cal A}[f,f^{\star}]=\int f^{\star\alpha}F_{\alpha}dx$ to be stationary while computing the variation with respect to the fields~$f$.

Ibragimov also discusses the generalized conservation theorem applicable to Lagrangians of arbitrary order. For fields that satisfy the Euler-Lagrange equation, $\delta{\cal L}/\delta u^{\alpha}=0$, the general theorem can be written as
\begin{equation}
\label{eq:fundamental_identity}
D_{\mu}j^{\mu}=X({\cal L})+{\cal L}D_{\mu}(\xi^{\mu}),
\end{equation}
and, in case of second order Lagrangians, the flux $j^{\mu}$ is
\begin{equation}
\label{eq:second_order_flux}
j^{\mu}=\xi^{\mu}{\cal L}+W^{\alpha}\frac{\partial{\cal L}}{\partial u^{\alpha}_{\mu}}+D_{\nu}(W^{\alpha})\frac{\partial{\cal L}}{\partial u^{\alpha}_{\mu\nu}}-W^{\alpha}D_{\nu}\left(\frac{\partial{\cal L}}{\partial u^{\alpha}_{\mu\nu}}\right),
\end{equation}
with $W^{\alpha}=\eta^{\alpha}-\xi^{\sigma}u_{\sigma}^{\alpha}$.

{\bf {\em Energy-momentum stress tensor}} --
Before attending the Rosenbluth-Fokker-Planck system, explicit conservation laws for a second order Lagrangian are derived. They are obtained by investigating phase-space-time symmetries, and by generalizing the classic definition of the energy-momentum stress tensor. 

Let us define an infinitesimal transformation with the generating vector field $X=C^{\mu}\partial/\partial x^{\mu}$ where $C^{\mu}$ are constants (constant phase-space-time translations). The flux $j^{\mu}$ defined in Eq.~(\ref{eq:second_order_flux}) then becomes $j^{\mu}=C^{\nu}T^{\mu\nu}$ where the generalized {\em energy-momentum stress tensor}, $T^{\mu\nu}$, applicable to second order Lagrangians has the explicit form
\begin{equation}
T^{\mu\nu}\equiv \delta^{\mu\nu}{\cal L}-u_{\nu}^{\alpha}\frac{\partial{\cal L}}{\partial u_{\mu}^{\alpha}}+u^{\alpha}_{\nu}D_{\sigma}\left(\frac{\partial{\cal L}}{\partial u_{\mu\sigma}^{\alpha}}\right)-u_{\nu\sigma}^{\alpha}\frac{\partial{\cal L}}{\partial u_{\mu\sigma}^{\alpha}}.
\end{equation}
 It is clear that if the Lagrangian is independent of second order derivatives, our definition reduces to $T^{\mu\nu}=\delta^{\mu\nu}{\cal L}-u_{\nu}^{\alpha}\partial{\cal L}/\partial u_{\mu}^{\alpha}$, i.e., to the standard energy-momentum stress tensor for first order Lagrangians.

We then apply Eq.~(\ref{eq:fundamental_identity}) to find $C^{\nu}D_{\mu}T^{\mu\nu}=C^{\nu}{\cal L}_{\nu}$ where the derivative of the Lagrangian with respect to coordinate $x^{\nu}$ is understood as an explicit derivative with respect to the independent variables. Since the $C^{\nu}$ are arbitrary, the equation holds component-wise, and we finally obtain
\begin{equation}
D_{\mu}T^{\mu\nu}={\cal L}_{\nu},\qquad \nu=0,\dots,n.
\end{equation}

It is then natural to consider the components of the energy-momentum stress tensor separately, and to define the {\em energy density}
\begin{equation}
\label{eq:energy-density}
{\cal E}\equiv{\cal L}-u_{t}^{\alpha}\frac{\partial{\cal L}}{\partial u_{t}^{\alpha}}+u_{t}^{\alpha}D_{\sigma}\left(\frac{\partial{\cal L}}{\partial u_{t\sigma}^{\alpha}}\right)-u_{t\sigma}^{\alpha}\frac{\partial{\cal L}}{\partial u_{t\sigma}^{\alpha}},
\end{equation}
the {\em energy-density flux}
\begin{equation}
\label{eq:energy-density-flux}
S^{i}\equiv-u_{t}^{\alpha}\frac{\partial{\cal L}}{\partial u_{i}^{\alpha}}\\+u^{\alpha}_{t}D_{\sigma}\left(\frac{\partial{\cal L}}{\partial u_{i\sigma}^{\alpha}}\right)-u_{t\sigma}^{\alpha}\frac{\partial{\cal L}}{\partial u_{i\sigma}^{\alpha}},
\end{equation}
the {\em momentum density}
\begin{equation}
\label{eq:momentum-density}
P^{i}=-u_{i}^{\alpha}\frac{\partial{\cal L}}{\partial u_{t}^{\alpha}}\\+u^{\alpha}_{i}D_{\sigma}\left(\frac{\partial{\cal L}}{\partial u_{t\sigma}^{\alpha}}\right)-u_{i\sigma}^{\alpha}\frac{\partial{\cal L}}{\partial u_{t\sigma}^{\alpha}},
\end{equation}
and the {\em stress tensor}
\begin{equation}
\label{eq:stress-tensor}
\Pi^{ij}\equiv \delta^{i}_{j}{\cal L}-u_{j}^{\alpha}\frac{\partial{\cal L}}{\partial u_{i}^{\alpha}}\\+u^{\alpha}_{j}D_{\sigma}\left(\frac{\partial{\cal L}}{\partial u_{i\sigma}^{\alpha}}\right)-u_{j\sigma}^{\alpha}\frac{\partial{\cal L}}{\partial u_{i\sigma}^{\alpha}},
\end{equation}
in terms of which, we write equations for the energy and momentum explicitly
\begin{align}
D_t{\cal E}+D_iS^{i}=&{\cal L}_t,\\
D_tP^j+D_i\Pi^{ij}=&{\cal L}_j.
\end{align}
It is clear that a conservation law exists for the energy (momentum) related to the fields $u^{\alpha}$ if the Lagrangian does not explicitly depend on time (phase-space coordinates). 

{\bf {\em Rosenbluth-Fokker-Planck system}} -- 
Instead of the integro-differential formulation by Landau, Rosenbluth found a way to formulate the collisional evolution of the distribution function in terms of a coupled system of partial differential equations~\cite{rosenbluth:PhysRev.107.1}. Here, for the sake of clarity, we consider only the single species system, which is normalized to dimensionless units. Including more species is straight-forward. The phase-space coordinates $x^i$ now denote the three Cartesian velocity coordinates. 

The non-linear Rosenbluth-Fokker-Planck system for the distribution function $f$ and for the potentials $\phi$ and $\psi$ is then defined by the differential equations
\begin{align}
F_{f}&\equiv\partial_t f-D_i\left(D_i\phi\, f-D_iD_j\psi\,D_jf\right),\\
F_{\phi}&\equiv D_iD_i\phi-f,\\
F_{\psi}^{(1)}&\equiv D_iD_i\psi-\phi,\\
F_{\psi}^{(2)}&\equiv D_iD_iD_jD_j\psi-f,
\end{align}
where we have denoted two different options for the equation $F_{\psi}=0$. We then simply introduce the auxiliary fields $(f^{\star},\phi^{\star},\psi^{\star})$ and obtain the Lagrangians
\begin{equation}
{\cal L}^{(i)}_{\textrm{RFP}}=f^{\star}F_{f}+\phi^{\star}F_{\phi}+\psi^{\star}F_{\psi}^{(i)}.
\end{equation}
As such, ${\cal L}^{(1)}_{\textrm{RFP}}$ contains derivatives of order three and ${\cal L}^{(2)}_{\textrm{RFP}}$ derivatives of order four. Addition of total derivatives into the Lagrangian does not, however, alter the Euler-Lagrange equations, and we may reduce the order of derivatives present in the Lagrangians.

Neglecting total derivatives, we obtain the second-order Lagrangians
\begin{align}
{\cal L}^{(1)}_{\textrm{RFP}}=&\,\frac{1}{2}f^{\star}\left(D_t-D_i\phi\,D_i\right)f-\frac{1}{2}f\left(D_t-D_i\phi\,D_i\right)f^{\star}\notag\\&-\frac{1}{2}ff^{\star}D_iD_i\phi-D_if^{\star}\,D_iD_j\psi\,D_jf\notag\\&-D_i\phi^{\star}D_i\phi-D_i\psi^{\star}D_i\psi-\phi^{\star}f-\psi^{\star}\phi,
\end{align}
\begin{align}
{\cal L}^{(2)}_{\textrm{RFP}}=&\,\frac{1}{2}f^{\star}\left(D_t-D_i\phi\,D_i\right)f-\frac{1}{2}f\left(D_t-D_i\phi\,D_i\right)f^{\star}\notag\\&-\frac{1}{2}ff^{\star}D_iD_i\phi-D_if^{\star}\,D_iD_j\psi\,D_jf\notag\\&-D_i\phi^{\star}D_i\phi+D_iD_i\psi^{\star}D_jD_j\psi-\phi^{\star}f-\psi^{\star}f,
\end{align}
and it is straight-forward to verify that both lead to the Rosenbluth-Fokker-Planck system when varied with respect to the fields $(f^{\star},\phi^{\star},\psi^{\star})$.

Neither ${\cal L}^{(1)}_{\textrm{RFP}}$ nor ${\cal L}^{(2)}_{\textrm{RFP}}$ depends explicitly on time $t$ or velocity $x^i$. It is thus possible to find exact conservation laws in the field-theory sense simply by computing the energy density ${\cal E}$, the energy-density flux $S^i$, the momentum density $P^i$, and the stress tensor $\Pi^{ij}$ according to their definitions given in equations~(\ref{eq:energy-density}--\ref{eq:stress-tensor}).

Further, since the conservation laws are expressed in divergence form and all our fields vanish at the boundaries of the infinite velocity space, we immediately obtain conservation of the total field-related energy
\begin{equation}
D_t\int{\cal E}\,dx^1dx^2dx^3=0,
\end{equation}
and component-wise conservation of the total field-related momentum 
\begin{equation}
D_t\int P^i\,dx^1dx^2dx^3=0, \quad i=1,2,3,
\end{equation}
where the explicit expressions for the energy density and momentum density are
\begin{align}
{\cal E}&={\cal L}+\frac{1}{2}fD_tf^{\star}-\frac{1}{2}f^{\star}D_tf,\\
P^i&=\frac{1}{2}fD_if^{\star}-\frac{1}{2}f^{\star}D_if.
\end{align} 

Although the definitions for energy and momentum carried by the extended Rosenbluth-Fokker-Planck system are not the physical kinetic energy and momentum carried by the distribution function $f$, the exact conservation laws in our extended continuous system motivate finding variational integrators that would provide discrete counterparts for the continuous conservation laws, and thus solve the extended system as well as possible. As the physical system can be derived from the extended one, it is anticipated that the superior properties of variational integrators, though applied to the extended system, would then carry to the physical system as well.

{\bf {\em Adjoint system and Langevin equation}} --
The adjoint equations for the fields $(f^{\star},\phi^{\star},\psi^{\star})$ are derived by applying the second order Euler-Lagrange operator
\begin{equation}
\frac{\partial{\cal L}}{\partial u^{\alpha}}-D_{\mu}\left(\frac{\partial{\cal L}}{\partial u^{\alpha}_{\mu}}\right)+D_{\mu}D_{\nu}\left(\frac{\partial{\cal L}}{\partial u^{\alpha}_{\mu\nu}}\right)\equiv F^{\star}_{\alpha},
\end{equation}
with respect to the fields $(f,\phi,\psi)$. For the Lagrangian ${\cal L}^{(1)}_{\textrm{RFP}}$ one obtains
\begin{align}
F^{\star,(1)}_{f}\equiv &\,-D_tf^{\star}+(D_i\phi)(D_if^{\star})\notag\\&+D_i\left(D_iD_j\psi D_jf^{\star}\right)-\phi^{\star},\\
F^{\star,(1)}_{\phi}\equiv &\,D_iD_i\phi^{\star}-\psi^{\star}-D_i\left(fD_if^{\star}\right),\\
F^{\star,(1)}_{\psi}\equiv &\,D_iD_i\psi^{\star}-D_iD_j\left(D_ifD_jf^{\star}\right),
\end{align}
and for the alternative Lagrangian ${\cal L}^{(2)}_{\textrm{RFP}}$ the computation gives
\begin{align}
F^{\star,(2)}_{f}\equiv &\,-D_tf^{\star}+(D_i\phi)(D_if^{\star})\notag\\&+D_i\left(D_iD_j\psi D_jf^{\star}\right)-\phi^{\star}-\psi^{\star},\\
F^{\star,(2)}_{\phi}\equiv &\,D_iD_i\phi^{\star}-D_i\left(fD_if^{\star}\right),\\
F^{\star,(2)}_{\psi}\equiv &\,D_iD_iD_jD_j\psi^{\star}-D_iD_j\left(D_ifD_jf^{\star}\right).
\end{align}

Considering then the case of test-particles in a given background plasma, the Rosenbluth potentials are pre-defined and no longer dependent variables: the Lagrangian becomes simply ${\cal L}_{\textrm{RFP}}=f^{\star}F_{f}$. Noting that $D_jD_j\psi=\phi$ we define the coefficients
\begin{align}
\mu^i&=-2D_i\phi\\
 \frac{1}{2}\sigma^{ik}\sigma^{jk}&=-D_iD_j\psi,
\end{align}
and write the equations for $f$ and $f^{\star}$ as
\begin{align}
F_{f}&=-D_tf-D_i\left(\mu^if\right)+D_iD_j\left(\frac{1}{2}\sigma^{ik}\sigma^{jk}\,f\right),\\
F^{\star}_{f}&=D_tf^{\star}+\mu^i\left(D_if^{\star}\right)+\frac{1}{2}\sigma^{ik}\sigma^{jk}\left(D_iD_jf^{\star}\right),
\end{align}
which are the Kolmogorov forward ($F_f=0$) and backward ($F^{\star}_{f}=0$) equations. The forward equation describes time evolution of the probability density of a stochastic process $X$ that obeys the Langevin equation of the It\'o type
\begin{equation}
dX^i=\mu^i(X)dt+\sigma^{ij}(X)dW^j,
\end{equation}
where $dW^j$ is an infinitesimal change in a Wiener process $W^j$ with expectation value $\mathrm{E}[W^j]=0$ and variance $\mathrm{Var}[W^j]=t$. The backward equation, on the other hand, describes the evolution of the probability density for $t\le s$ given end condition $f^{\star}(x)=p(x)$.

{\bf {\em Summary}} --
In this letter, we have derived an Eulerian action principle for Coulomb collisions in plasmas using the Rosenbluth potentials and the concept of extended Lagrangians. We also derived the exact conservation laws for the total energy and momentum carried by the fields in the extended system. In the case of a test-particle population in a given background plasma, the action principle was shown to equal the Langevin equation for individual particles.

By providing an action principle for statistical description of Coulomb collisions, this letter, first of all, enables for a comprehensive field theory in plasma physics. Secondly, as the action given is in a form suitable for discretization, the letter facilitates the development of variational integrators for the Fokker-Planck collision operator. Finally, derivation of exact conservation laws allows variational algorithms to be benchmarked and validated.

The author is grateful for the encouraging comments from T.~F\"ul\"op, J.~Candy, I.~Pusztai, A.~Stahl, O.~Emb\'eus, G.~Papp, A.~Bhattacharjee, T.~Kurki-Suonio, S.~Newton, and A.~Brizard and for all the fruitful discussions.

\bibliographystyle{unsrt}
\bibliography{bibfile}

\end{document}